\documentclass[11pt,twoside]{article}
\usepackage{./asp2014}

\aspSuppressVolSlug
\resetcounters

\begin{document}

\title{Detecting Diffuse Sources in Astronomical Images}
\author{T. Butler-Yeoman,$^1$ M. Frean,$^1$ C.P. Hollitt,$^1$ D.W. Hogg,$^2$ and M. Johnston-Hollitt$^3$
\affil{$^1$School of Engineering \& Computer Science, Victoria University of Wellington, PO Box 600, Wellington 6140, New Zealand; \email{Marcus.Frean@vuw.ac.nz}}
\affil{$^2$Center for Cosmology \& Particle Physics, Department of Physics, New York University, 4 Washington Place \#424, New York , NY 10003, USA;}
\affil{$^3$School of Chemical \& Physical Sciences, Victoria University of Wellington, PO Box 600, Wellington 6140, New Zealand}}

% This section is for ADS Processing.  There must be one line per author.
\paperauthor{T. Butler-Yeoman}{Author1Email@email.edu}{}{Victoria University of Wellington}{School of Engineering and Computer Science}{Wellington}{Wellington}{6140}{New Zealand}
\paperauthor{M. Frean}{Marcus.Frean@vuw.ac.nz}{}{Victoria University of Wellington}{School of Engineering and Computer Science}{Wellington}{Wellington}{6140}{New Zealand}
\paperauthor{C.P. Hollitt}{Christopher.Hollitt@vuw.ac.nz}{}{Victoria University of Wellington}{School of Engineering and Computer Science}{Wellington}{Wellington}{6140}{New Zealand}
\paperauthor{D.W. Hogg}{david.hogg@nyu.edu}{}{New York University}{Department of Physics}{New York}{New York}{NY 10003}{USA}
\paperauthor{M. Johnston-Hollitt}{Melanie.Johnston-Hollitt@vuw.ac.nz}{}{Victoria University of Wellington}{School of Chemical and Physical Sciences}{Wellington}{Wellington}{6140}{New Zealand}

\begin{abstract}
We present an algorithm capable of detecting diffuse, dim sources of any size in an astronomical image. These sources often defeat traditional methods for source finding, which expand regions around points of high intensity. Extended sources often have no bright points and are only detectable when viewed as a whole, so a more sophisticated approach is required. Our algorithm operates at all scales simultaneously by considering a tree of nested candidate bounding boxes, and inverts a hierarchical Bayesian generative model to obtain the probability of sources existing at given locations and sizes. This model naturally accommodates the detection of nested sources, and no prior knowledge of the distribution of a source, or even the background, is required. The algorithm scales nearly linear with the number of pixels making it feasible to run on large images, and requires minimal parameter tweaking to be effective. We demonstrate the algorithm on several types of astronomical and artificial images. %, and show that (on tested data) the algorithm can detect most significant sources, with a low rate of false positives. 
\end{abstract}

\section{Introduction}
With a number of new telescope facilities currently in the construction or design phases, such as the Square Kilometre Array which is expected to produce exabytes of data, there is a pressing need to develop statistically robust detection and classification algorithms which can detect all astronomical sources of interest, including faint and extended emission. Detection and characterisation of objects in astronomical images has been examined extensively for unresolved point sources (e.g. \cite{Hopkins15} and references therein) and bright extended sources which lie above a threshold \citep{Whiting12,Hancock12}, but few algorithms have been deployed which are capable of detecting extended regions of  low surface brightness. Detection of faint interesting sources, such as the lobes of radio galaxies, has largely been an citizen science exercise \citep{Banfield15}, though some novel work has recently started \citep{Hollitt12,Frean14}. 

We introduce Oddity, a detection algorithm that outputs boxes around sources. %, and demonstrate it on constructed and real data. %An example of its output is shown in figure \ref{f:intro_image}.
Oddity is based on a tree-based generative model of an image in which box-shaped regions of the sky have intensity distributions (after discretization) that are \emph{anomalous relative to their surroundings}. The algorithm finds sources via a tractable Bayesian inversion of this model. The background distribution is obtained as a by-product.

\section{A tree of boxes: Source-finding by inversion of a simple generative model} 
Each source is modelled by a Dirichlet Compound Multinomial (DCM) distribution as follows. First, a single categorical $\vec h$ is drawn from a Dirichlet distribution having hyperparameters $\vec \alpha$. Then, the intensity value of each pixel belonging to that source is drawn i.i.d from the categorical (see \citet{friedlander,Frean14}), leading to bin counts $\vec x$. 
The DCM allows observations to be modelled as drawn from an \emph{unknown} multinomial distribution. We set the hyperparameters $\vec \alpha$ to $\vec 1$ in all cases here, which corresponds to assuming \emph{no prior knowledge} about the intensity profile of any source or the background. 
The DCM is unusual among compound distributions in that the likelihood $P_\text{DCM}(\vec x \mid \vec \alpha)$ is readily evaluated.
For simplicity we take the regions of interest to be simple bounding boxes. This is a generally recognised format by which a source location can be returned to an astronomer, and conveniently affords rapid computation of the relevant bin counts $\vec x$ via a technique borrowed from computer graphics (\cite{crow1984summed}).
The image is initially modelled by a 4-ary tree of nodes, with the root node being the whole image. Children correspond to the parent's box split evenly into 4, and this tree continues down as far as needed. % (and could go as far as individual pixels).
Each node $S_i$ can be in two states, either \emph{active} or \emph{inactive}. If $S_i$ is active, it is a source, and the pixels $x^{(i)}$ in its box are generated by a DCM specific to $S_i$. However if some node $S_j$ in $S_i$'s subtree is also active, then the pixels in $S_j$'s box are taken to be generated from $S_{j}$, not $S_i$, and so active nodes take precedence over their active parents. Under this model, the `background' of the image is simply a source at the root of the tree and $S_0$ is defined to always be active. For a given assignment of states, the log likelihood for the whole image is simply the sum of $P_\text{DCM}(\vec x^{(i)} \mid \vec \alpha)$ for all active nodes.

\begin{figure}[]
  \centering
\begin{tabular}{ccc}
 {\includegraphics[width=4cm]{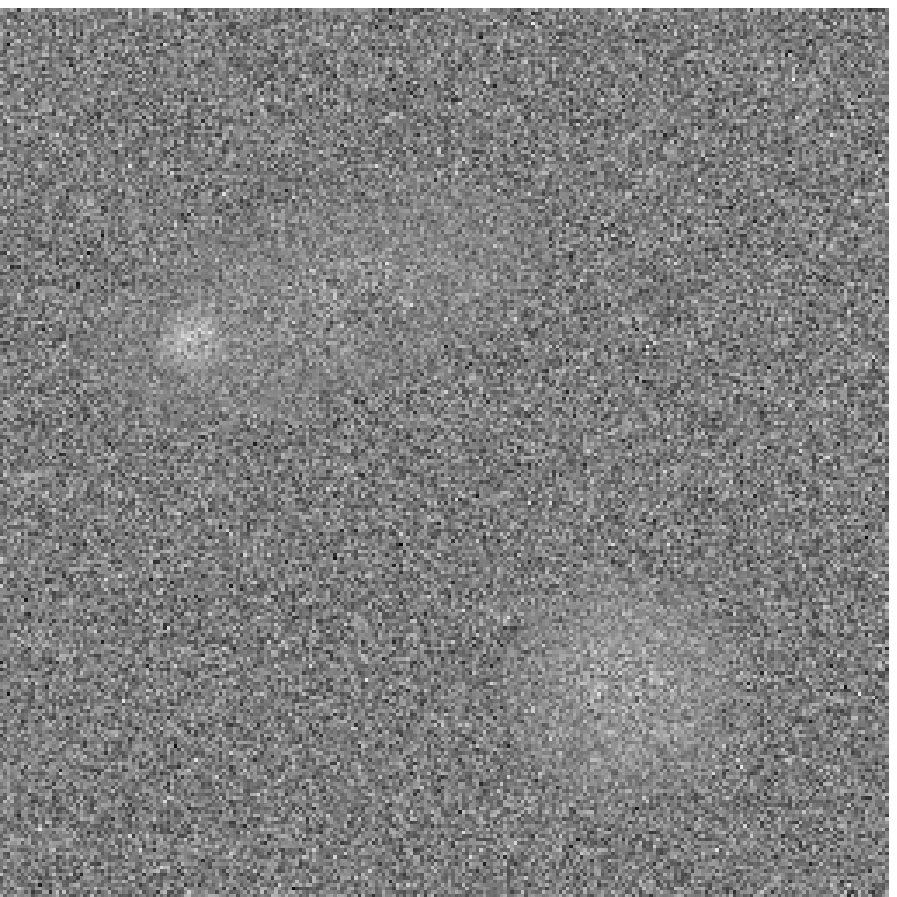}}
 & {\includegraphics[width=4cm]{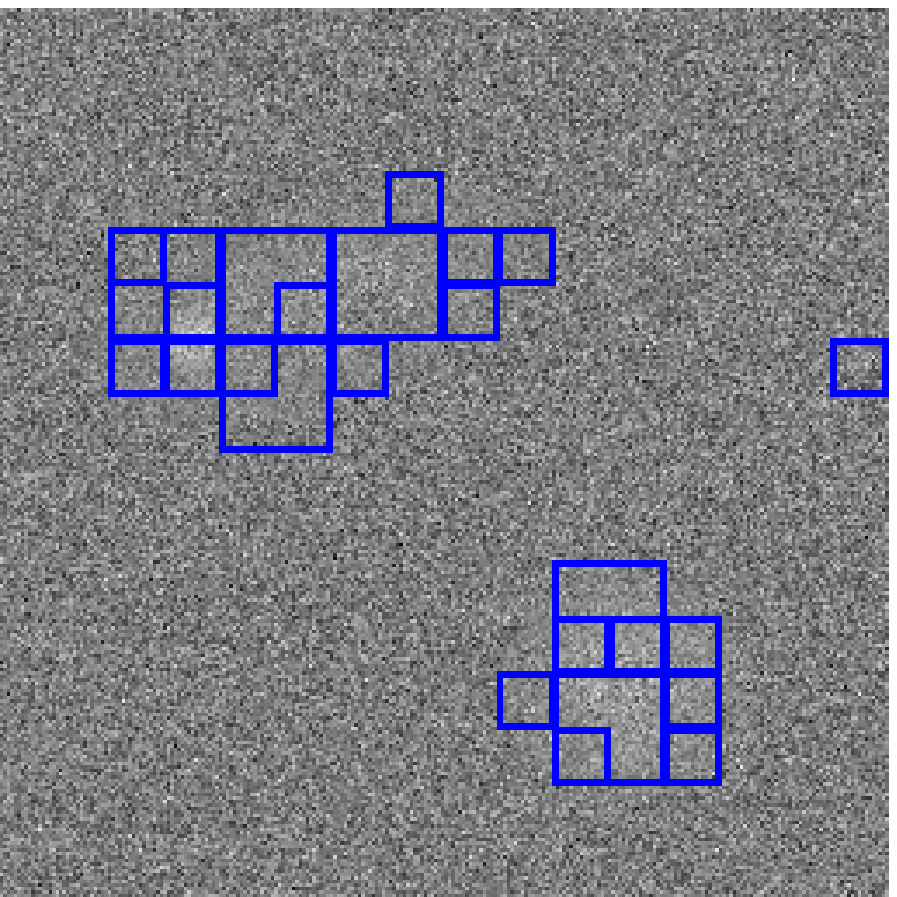}}
 & {\includegraphics[width=4cm]{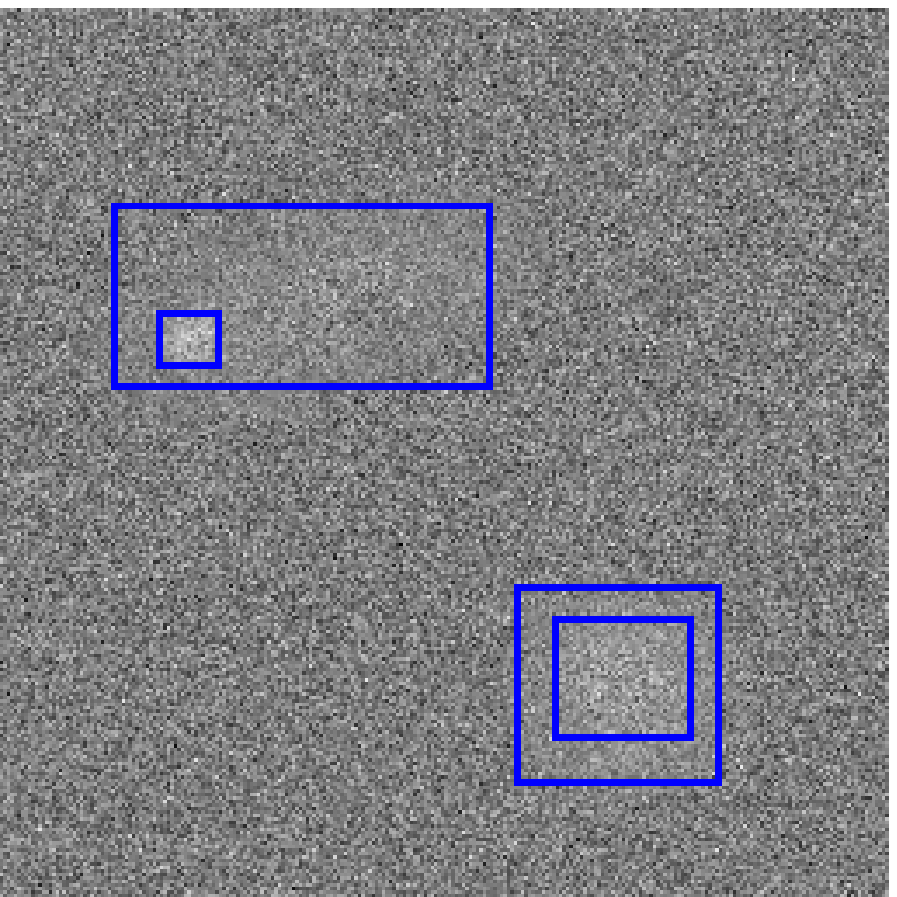}}  \\
 (a) Raw image & (b) Oddity stage 1 & (c) Oddity final \\\\ 
 {\includegraphics[width=4cm]{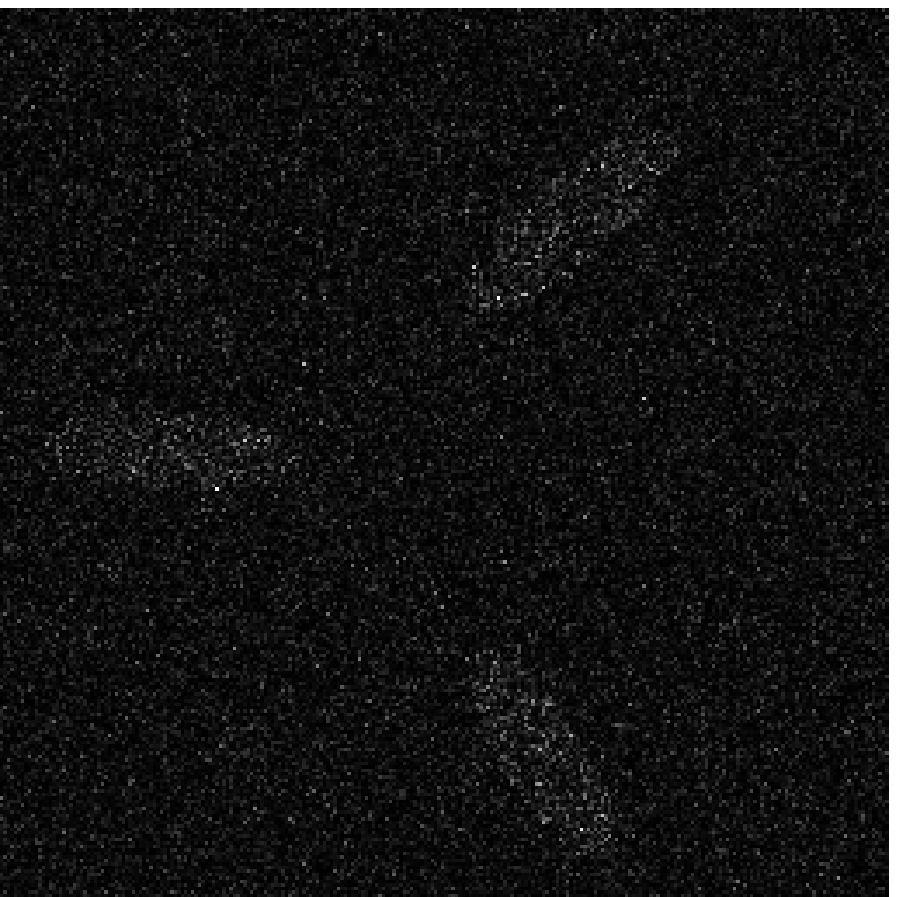}}
 & {\includegraphics[width=4cm]{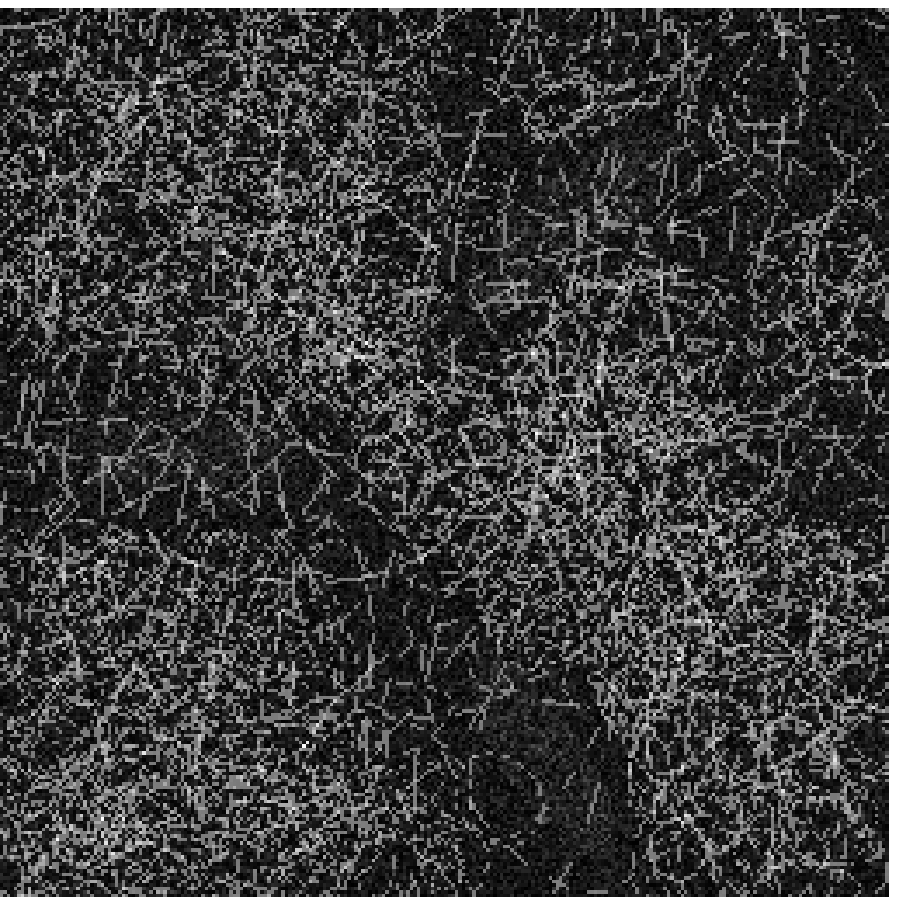}}
 & {\includegraphics[width=4cm]{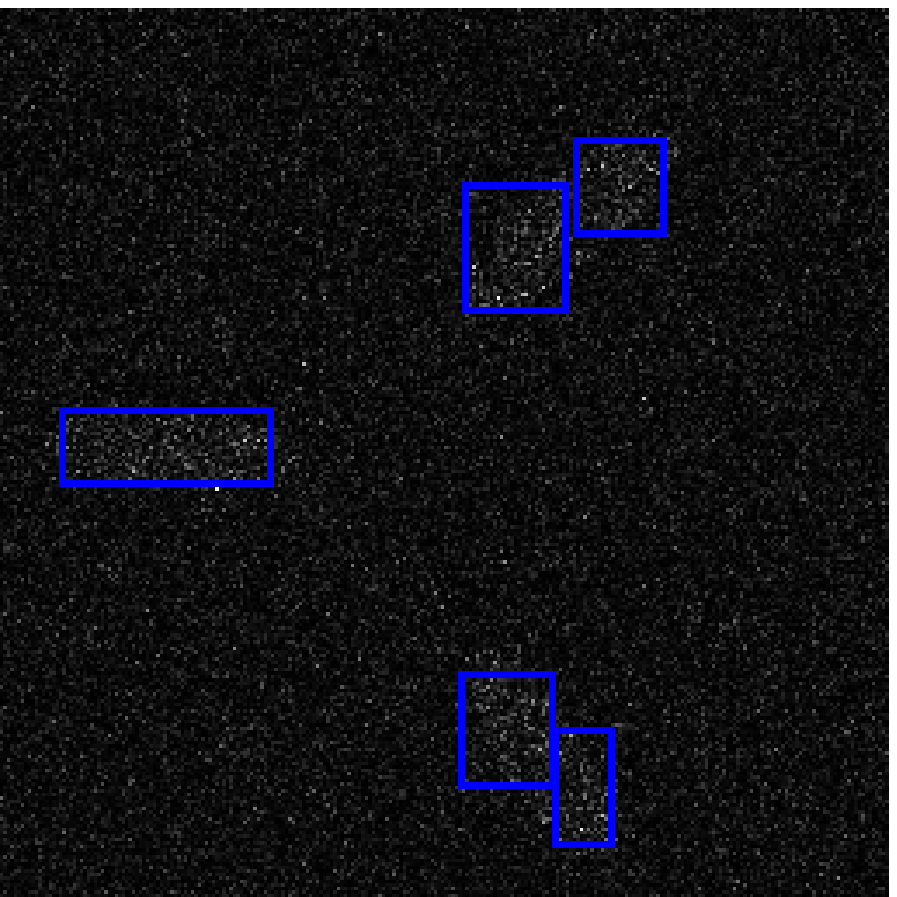}}  \\
 (d) Exponential & (e) SExtractor & (f) Oddity \\\\
 {\includegraphics[width=4cm]{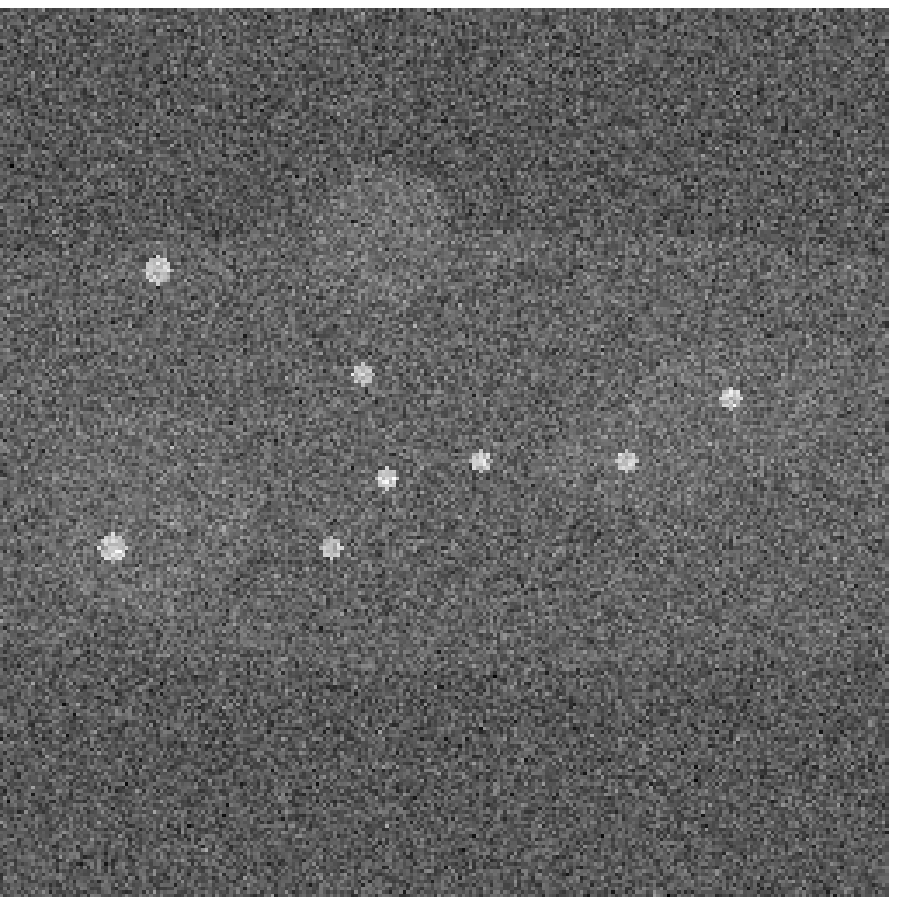}}
 & {\includegraphics[width=4cm]{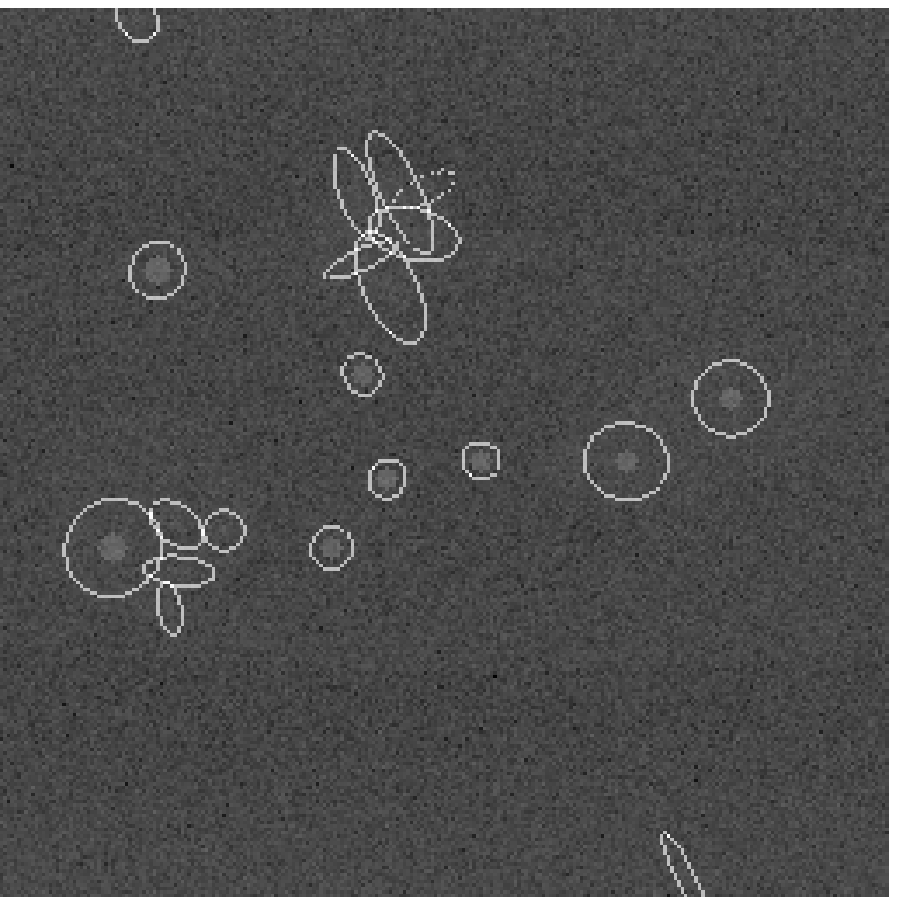}}
 & {\includegraphics[width=4cm]{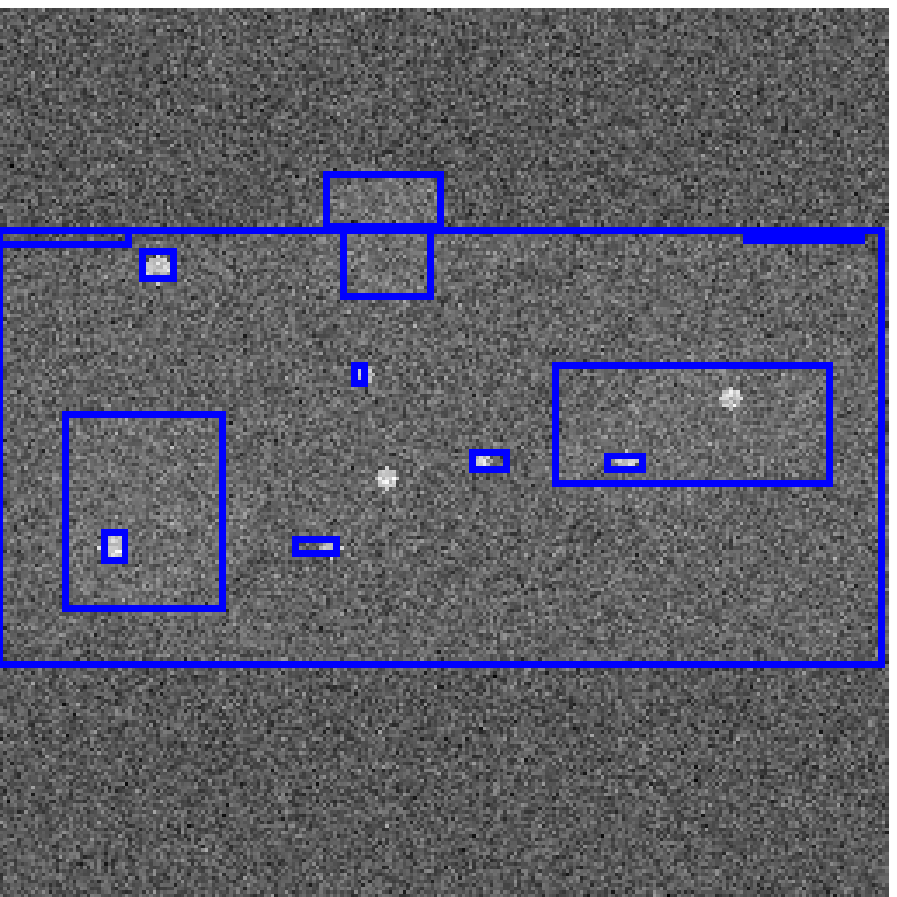}}
\\
 (g) Nested sources & (h) SExtractor & (i) Oddity \\\\
\end{tabular}
{\includegraphics[width=8.6cm]{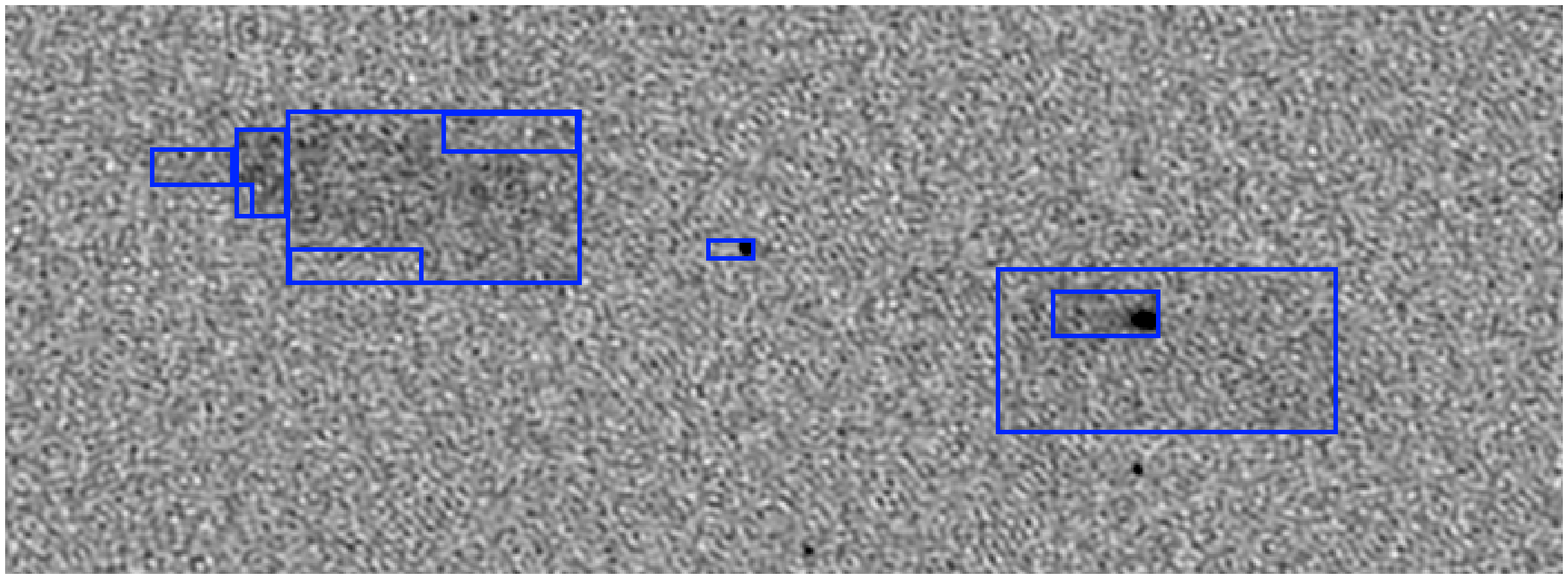}}

  \caption[Examples of Oddity running on a constructed images]{{\it Top row:} a constructed image, with pixel intensities that are Gaussian. (b) shows the result of the first stage, and (c) after the second. {\it Second row:} Output of SExtractor (middle) and Oddity (right) on an image (left) having an exponential distribution for pixel intensities. {\it Third row:} Outputs for an image with complex nested sources. {\it Bottom row:} Detecting a faint radio galaxy in the ATLBS survey (see text). }  
  \label{f:various_examples}
\end{figure}

Given a generative model, source finding becomes the task of inverting that model to infer plausible joint states $\vec S$. We achieve this in two stages. First, Gibbs Sampling is used to exhibit configurations that describe the data well. The sampler visits all nodes multiple times and switches $S_i$ to be active/inactive with probability given by a straightforward Gibbs update rule derived from the log likelihood. We initialise the state of all nodes to be inactive, except the root which is always active. A modest number of burn-in iterations are performed to allow the Markov chain to mix before drawing samples (with a gap of a few iterations between samples). We found 20, 10 and 2 for these numbers is sufficient, indicating that the chain has a short mixing time in most cases.
%Figure \ref{f:gibbs_1D} shows a toy example in 1D for illustration.
Any box that was active in more than some proportion $T$ of the samples so obtained can now be accepted as a possible source, and a second stage of refinement begins in which the severe constraints imposed on the model for tractability are relaxed. The second stage is an optimization rather than a sampler: it ``fine-tunes'' the positions of sources, merges them, or removes them altogether. The quantity being optimized is  exactly the same as that used in the first stage, namely the log likelihood of the entire image under the tree-based DCM generative model. The structure now becomes a general tree and the box positions are no longer constrained, apart from the requirement that they remain entirely enclosed by their parent box. The latter is a condition that appears important in order for the algorithm to scale to large images.

\section{Performance}
Figure \ref{f:various_examples} shows several examples, illustrating the algorithm's robustness to various aspects of image statistics. We first discretize each image, by setting thresholds such that bins achieve approximately equal occupancy. The output of SExtractor \citep{bertin1996sextractor} (Source Extractor) is shown for comparison. While SExtractor is primarily used to find high surface brightness objects, it has parameter settings designed to be applicable to low surface brightness (LSB) objects. 
A recent and comprehensive study by \citet{Hopkins15} has shown that it remains competitive with newer techniques on both point and extended source extraction problems. %We are currently pursuing comparisons with other well regarded source detection algorithms.

Figure \ref{f:various_examples}(g) is a rough simulation of the `galactic plane' (the large, dim band across the image). Oddity successfully separates this out, as well as the three large but dim sources. The top-most large source is detected as two separate boxes because it partially overlaps the galactic plane, consistent with the restrictions made. Six of the eight point sources are detected, while the other two are not. This is not a particular concern as they are not the target of the algorithm, but performance could be improved on these types of sources by using a more sophisticated binning scheme. %Note that Oddity returns two `overdetection' boxes at the corners of the galactic plane (the one in the top-left is very slim). SExtractor detects all eight points sources, but misses the dim sources entirely.

The last row in the figure shows an example on real data from the Australia Telescope Low-Brightness Survey (ATLBS) \citep{subrahmanyan2010atlbs}. ATLBS was a survey of two regions of the sky that produced very high resolution images with almost no imaging artefacts. These images are considered good models of the eventual output of the SKA and its precursors, and were created in part to be a test for new source detection techniques. Note the correct identification of the nested source in this image.

Our algorithm has a worst case cost that is $\mathcal{O}(n \log n)$ in the number of pixels and quadratic in the number of active nodes. Our current implementation typically runs in under 1 second for $n=10^6$ pixels and exhibits near-linear scaling with $n$, unless the density of sources is very high.

\acknowledgements MF, CPH \& MJ-H are supported by the New Zealand Ministry of Business, Innovation and Employment under an SKA pre-construction grant.

\bibliography{O2-4} 

\begin{thebibliography}{}
\expandafter\ifx\csname natexlab\endcsname\relax\def\natexlab#1{#1}\fi
\expandafter\ifx\csname url\endcsname\relax
  \def\url#1{\texttt{#1}}\fi
\expandafter\ifx\csname urlprefix\endcsname\relax\def\urlprefix{URL }\fi
\providecommand{\eprint}[2][]{\url{#2}}

\bibitem[{{Banfield} et~al.(2015){Banfield}, {Wong}, {Willett}
  et~al.}]{Banfield15}
{Banfield}, J.~K., {Wong}, O.~I., {Willett}, K.~W., et~al. 2015, \mnras, 453,
  2326. \eprint{1507.07272}

\bibitem[{Bertin \& Arnouts(1996)}]{bertin1996sextractor}
Bertin, E., \& Arnouts, S. 1996, Astronomy and Astrophysics Supplement Series,
  117, 393

\bibitem[{Crow(1984)}]{crow1984summed}
Crow, F.~C. 1984, ACM SIGGRAPH computer graphics, 18, 207

\bibitem[{{Frean} et~al.(2014){Frean}, {Friedlander}, {Johnston-Hollitt}, \&
  {Hollitt}}]{Frean14}
{Frean}, M., {Friedlander}, A., {Johnston-Hollitt}, M., \& {Hollitt}, C. 2014,
  Bayesian Inference and Maximum Entropy Methods in Science and Engineering,
  1636, 55

\bibitem[{{Friedlander}(2014)}]{friedlander}
{Friedlander}, A.~M. 2014, Master's thesis, Victoria University of Wellington,
  New Zealand

\bibitem[{{Hancock} et~al.(2012){Hancock}, {Murphy}, {Gaensler}, {Hopkins}, \&
  {Curran}}]{Hancock12}
{Hancock}, P.~J., {Murphy}, T., {Gaensler}, B.~M., {Hopkins}, A., \& {Curran},
  J.~R. 2012, \mnras, 422, 1812. \eprint{1202.4500}

\bibitem[{{Hollitt} \& {Johnston-Hollitt}(2012)}]{Hollitt12}
{Hollitt}, C., \& {Johnston-Hollitt}, M. 2012, Publications of the Astronomical
  Society of Australia, 29, 309. \eprint{1204.0382}

\bibitem[{{Hopkins} et~al.(2015){Hopkins}, {Whiting}, {Seymour}
  et~al.}]{Hopkins15}
{Hopkins}, A.~M., {Whiting}, M.~T., {Seymour}, N., et~al. 2015, ArXiv e-prints.
  \eprint{1509.03931}

\bibitem[{Subrahmanyan et~al.(2010)Subrahmanyan, Ekers, Saripalli, \&
  Sadler}]{subrahmanyan2010atlbs}
Subrahmanyan, R., Ekers, R., Saripalli, L., \& Sadler, E. 2010, Monthly Notices
  of the Royal Astronomical Society, 402, 2792

\bibitem[{{Whiting}(2012)}]{Whiting12}
{Whiting}, M.~T. 2012, \mnras, 421, 3242. \eprint{1201.2710}

\end{thebibliography}

\end{document}